\documentclass[twocolumn,epsfig,graphics,mathbbm,nopacs,hyperref]{revtex4-1}
\usepackage{amsmath,amsfonts,amssymb,amsthm,graphics,graphicx,epsfig,times,bbm,verbatim}
\usepackage{tikz}

\usepackage[utf8]{inputenc}

\newtheorem{definition}{Definition}
\newtheorem{theorem}{Theorem}

\newcommand{\proofend}{\hfill\rule{2mm}{2mm} }

\usepackage{amsfonts}
\usepackage{psfrag}
\usepackage{dsfont}
\usepackage[dvips]{rotating}
\usepackage{epsfig}
\usepackage{pstricks,pst-grad}

\newcommand{\md}{\mathrm{d}}
\newcommand{\Tr}{\mathrm{Tr}}

\newcommand{\id}{\mathbbm{1}}  
\newcommand{\PP}{\mathbbm{P}}
\newcommand{\E}{\mathbbm{E}}
\newcommand{\eps}{\varepsilon}

\newcommand{\rr}{\mathbbm{R}}

\newcommand{\U}{U}
\newcommand{\SU}{S\U}
\newcommand{\newproof}{{\it Proof: }}

\begin{document}

\unitlength=1mm

\title{Efficient and feasible state tomography of quantum many-body systems}
 
\date{\today}
\author{M.\ Ohliger$^{1,2}$, V.\ Nesme$^{1,3}$, and J.\ Eisert$^{1}$}

\affiliation{1 Dahlem Center for Complex Quantum Systems, Freie Universit{\"a}t Berlin, 14195 Berlin, Germany}

\affiliation{2 Institute for Physics and Astronomy, University of Potsdam, 14476 Potsdam, Germany}

\affiliation{3 Laboratoire d'Informatique de Grenoble, 38400 Saint-Martin-d'H\`eres, France}

\begin{abstract}
We present a novel method to perform quantum state tomography for many-particle systems which are particularly
suitable for estimating states in lattice systems such as of ultra-cold atoms in optical lattices. 
We show that the need for measuring a tomographically complete set of observables can be overcome by letting the state evolve under some suitably chosen random circuits followed by the measurement of a single observable. We generalize known results about the approximation of unitary $2$-designs, i.e., certain classes of random unitary matrices, by random quantum circuits and connect our findings to the theory of quantum compressed sensing. We show that for ultra-cold atoms in optical lattices established experimental
techniques like optical super-lattices, laser speckles, and time-of-flight measurements are sufficient to perform fully certified, 
assumption-free tomography. This is possible without the need of addressing single sites in any step of the procedure. 
Combining our approach with tensor network methods -- in particular the theory of matrix-product states -- we identify situations where the 
effort of reconstruction is even constant in the number of lattice sites, allowing in principle
to perform tomography on large-scale systems readily 
available in present experiments.
\end{abstract}

\maketitle

\section{Introduction}
\label{sec:intro}
Quantum state tomography is -- for obvious reasons -- a procedure of great importance
in a large number of experiments involving quantum systems: It amounts to
reconstructing an unknown quantum state entirely based on experimental data.
In many situations one indeed aims at identifying what state has actually been 
prepared in an experiment. This seems particularly important in the context of 
quantum information science, where quantum state and process tomography is now routinely
applied to small, precisely controlled quantum systems \cite{tomography1,tomography2,tomography3}. 
Yet, needless
to say, in a number of other contexts the reliable reconstruction of quantum states is an important aim 
as well.

For finite-dimensional quantum systems, conventional quantum state tomography can be performed by choosing a suitable basis of $\mathcal{B}(\mathbbm{C}^d)$, i.e., the operators on the $d$-dimensional 
Hilbert space of the system in question.
Then, the expectation values of these $d^2$ observables are being measured to some required accuracy,
from which one can reconstruct the unknown density matrix $\rho$. The same approach, however, is doomed
to failure when applied to quantum many-body systems:
If one has a many-body system at hand 
with $k$ lattice sites of local dimension $d_l$, the number of necessary different measurement settings is 
given by
$m=d_l^{2k}$, i.e., it scales exponentially with the size, rendering the treatment even of reasonably large systems
impossible. Techniques of quantum compressed sensing \cite{candesTao,davidletter,cvcs,david,Kosut} 
allow to significantly reduce the required number of measurement settings, if the state is of rank $r$, to $m=\Theta(rd\log^2 d)$ (where $\Theta$ denotes asymptotic equality). 
If $r\ll d$, this is an impressive reduction, and gives rise to feasible quantum state tomography 
in medium-sized quantum systems, 
but this number is still exponential in the number of sites. Such a scaling cannot be overcome without further restriction of the class of possible states, simply 
because even a pure state needs of the order of $d$ parameters to be described. However, if the state is not only pure but also described by a generic matrix product state (MPS), the necessary number of measurements only scales linearly with the system size and is even independent of it for the important special case of translationally invariant MPS.
What is more, in several instances the classical procedure to reconstruct the MPS matrices from the measurement data is efficient \cite{mps}. 

This small number of parameters ought to make tomography an easier task, but in many practical settings involving quantum many-body systems, a serious challenge arises: In most interesting systems it is very difficult if not impossible
to directly measure a full operator basis.
Instead, merely measurements of some preferred observables might be readily available. 

\begin{figure}
\centering
\includegraphics[width=0.6\linewidth]{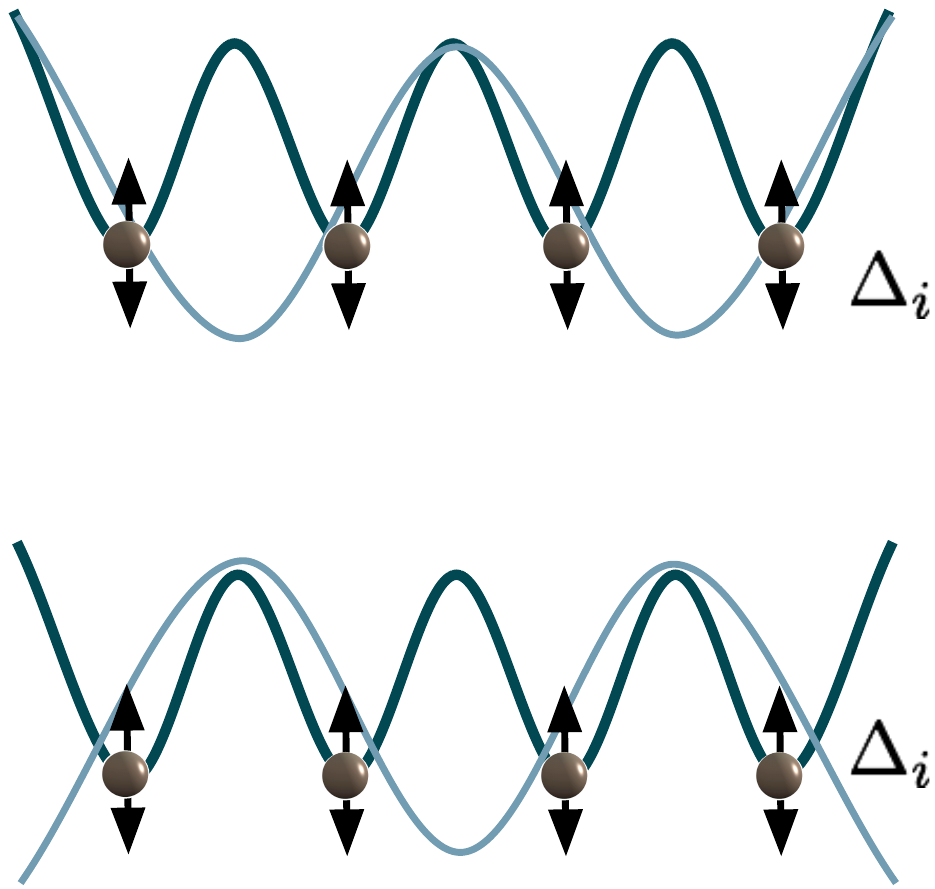}
\caption{\label{fig:optlat}Realization of a random quantum circuit by an optical superlattice. The lattice with the larger period is switched between the two depicted situations, and the lattice depth is changed locally by $\Delta_i$ which is uncorrelated between the lattice sites.}
\end{figure}

In the present work, we propose a solution to this problem. We do so by 
combining the action of a suitable random quantum circuit with a measurement of a very small
number of different observables (even a single one can be enough). Such random circuits
are just becoming a tool of great theoretical importance in several subfields 
of quantum information theory \cite{randomcircuits2,expspeedups,lowthesis}. 

Here, we show that they also offer significant technological 
advantages, in that they allow for the natural implementation of {\it complete quantum state tomography}
in systems of ultra-cold atoms using only the techniques of super-lattices and laser speckles. Both techniques have already been experimentally proven to be feasible \cite{superlattices,superlatticesrelax,speckle}. Combining these new insights with the above mentioned methods of MPS tomography brings, for the first time, full tomography of many-particle systems close to experimental reach. Note that once the state is 
estimated, one can from this knowledge of course also compute properties such as {\it entanglement entropies} of systems of cold atoms
\cite{Demler} -- but also all other properties that are a function of the quantum state.

Needless to say, it still amounts to a very difficult prescription.
But while conventional measurements in ultra-cold bosonic systems amount to measuring certain
correlation functions or estimates of the temperature in thermometry, say, the path described here eventually
allows for the full reconstruction of an unknown state of a quantum many-body system -- a very promising perspective.

The remainder of this article is organized as follows: First, we introduce in Section \ref{sec:tomtime} tomographically complete sets of observables and a generalization of operator bases, called tight frames, and discuss their realization by means of random unitary matrices. In a next step, we show in Section \ref{sec:undesign} how efficient compressed sensing is possible with families of unitary matrices which form approximate unitary $2$-designs
\cite{unidesign1, unidesign2, expspeedups, randomcircuits2, lowthesis, randomcircuits}. In Section \ref{sec:rand}, a way to efficiently realize such unitary $2$-designs with the help of random quantum circuits is presented, before we discuss the application of this approach to ultra-cold atoms in optical lattices in Section \ref{sec:optlat}. Finally, Section \ref{sec:mps} shows how, under certain assumptions on the state, tomography is possible with a number of measurements which is linear or constant in the number of lattice sites before a conclusion is given in Section \ref{sec:conc}. 

\section{Tomography by means of unitary evolution}
\label{sec:tomtime}
Before we turn to the question of how tomography is possible, we provide
general conditions for sets of observables to be suitable for reconstructing quantum states. We discuss why tomographic completeness as such 
is not sufficient and introduce generalizations of operator bases 
which allow for robust tomography.

\subsection{Tomographically complete sets of observables and tight frames}

We consider a quantum system consisting of $k$ subsystems, called (lattice) sites which all have local dimension $d_l$. Let $\rho\in {\cal S}({\cal H})$ be a quantum state on the Hilbert space ${\cal H}=\mathbbm{C}^d$ with dimension $d=d_l^k$ and $S$ the set of corresponding normalized observables, i.e., Hermitian matrices $w\in {\cal B}({\cal H})$ 
with $\|w\|_2=1$. In this whole work, we denote by $\|\cdot\|_p$ the Schatten $p$-norm, where $p=2$ and $p=\infty$ are the Frobenius norm and operator norm respectively. Furthermore, we use the Hilbert-Schmidt scalar product defined as $(A,B)=\Tr(A^\dag B)$ and the projection on the subspace spanned by some $w\in S$ which is defined as
\begin{equation}
\mathcal{P}_w:\rho\mapsto (w,\rho)w.
\end{equation}
Noting that a measurement of the expectation value of $w$ corresponds to determining $(w,\rho)$, we define the total sampling operator
\begin{equation}
\label{eq:defW}
\mathcal{W}_d=d^2\int\md\mu(w)\mathcal{P}_w\in\mathcal{B}(\mathcal{B}(\mathcal{H})),
\end{equation}
where $\mu$ is a probability measure on $S$.

A finite set of observables is said to be tomographically complete if any two different quantum states have distinct  expectation values for some observable:  This implies that one can theoretically reconstruct the state if one knows all expectation values. If $\mathcal{W}_d$ has full rank, i.e. ${\rm rank} (\mathcal{W}_d)=d^2$, the state $\rho$ can be obtained from $\mathcal{W}_d(\rho)$ by matrix inversion if issues of statistical errors and numerical imprecisions are neglected. For general probability measures we make use of the subsequent definition.

\begin{definition}[Tomographic completeness]
\label{def:tomcomplete}
The measure $\mu$ is said to be tomographically complete if  $\mathcal{W}_d$, as defined in Eq.\ (\ref{eq:defW}), is full rank.
\end{definition}

In practice, not every tomographically complete measure on the observables is necessarily useful: The observables, viewed as vectors in $\mathbbm{R}^{d^2}$, should not be too unevenly distributed over the sphere.  That is to say, if the ratio between the largest and smallest eigenvalues of $\mathcal{W}_d$ is large, small errors in the expectation values can lead to large errors in the reconstructed state.
The ideal situation is that of a tight frame, also known as spherical $1$-design:
\begin{definition}[Tight frame] 
\label{def:tightframe}
A probability measure $\mu$ on the set of 2-norm normalized Hermitian matrices $S$ is called a tight frame if $\mathcal{W}_d=\id$ with $\mathcal{W}_d$ given by Eq.\ (\ref{eq:defW}).
 \end{definition}
Examples for tight frames include any operator basis and the rotationally invariant measure on  the 2-norm sphere $S$ \cite{cvcs}. When observables are taken from tight frames, the reconstruction problem is well conditioned and small errors in the expectation values only lead to a small error in the reconstructed state \cite{ripless,cvcs}.

\subsection{Evolution of observables}
A tomographically complete set of observables must contain at least $d^2$ observables, which might be difficult to measure directly. We introduce a way to do tomography by performing suitably random unitaries, followed by the estimation of the expectation value for a single observable.  A different but related approach has been employed in Ref.\ \cite{randomtomography} to perform high-fidelity quantum state reconstruction in situations where the knowledge about the state is not tomographically complete. In the present work, the time evolution is a tool to obtain knowledge about the quantum state of the system leaving the question of determining Hamiltonians aside \cite{daniel}.

Switching to the Heisenberg picture, the outlined procedure amounts to measuring the time-evolved observables. Simple as this idea is, it allows the economical reconstruction of unknown quantum states, as it turns out:

\begin{definition}[Induced observables]
\label{def:inducedobs}
To a measure $\tilde{\mu}$ on the special unitary group $\SU(d)$ and an observable $w_0$ we associate the following induced measure on $S$:
\begin{equation}
\label{eq:defindmu}
	\mu=\biggl(1-\frac{1}{d^2}\biggr)\tilde{\mu}\circ f^{-1} +\frac{1}{d^2} \delta_{\id/\sqrt{d}}
\end{equation}
where $f:\U(d)\to S$ is defined by $f(U)=U^\dagger w_0 U$, and $\delta_x$ denotes the Dirac in $x$. Notice that if $\tilde{\mu}$ is a probability measure, then so is $\mu$.
\end{definition}
An important insight is provided by the following observation.
 
\begin{theorem}[Tight frame induced by random unitary]
\label{obs:inducedtf}
Let $w$ be a traceless, normalized observable and $\mu_{\rm H}$ be the Haar measure on $\SU(d)$. The measure induced on $S$ is a tight frame.
\end{theorem}

\newproof 
To show this observation, we use (\ref{eq:defindmu}) to calculate the sampling operator (\ref{eq:defW}) for a state $\rho$ with $\Tr(\rho)=1$:
\begin{equation}
\label{eq:obstemp0}
\mathcal{W}_{d,\rm H}(\rho)=(d^2-1)\int\md\mu_{\rm H}(U)(U^\dag w_0 U,\rho)U^\dag w_0 U+\frac{\id}{d}.
\end{equation}
As $w$ is diagonalized by a unitary matrix, we can assume it to be diagonal and obtain
\begin{align}
\left(\mathcal{W}_{d,\rm H}(\rho)\right)_{i,j}=&(d^2-1)\sum_{k,l,m,n}\E_{U\sim\mu_{\rm H}}\left[U_{k,l}U_{m,j}\bar{U}_{k,n}\bar{U}_{m,i}\right]\nonumber\\
&\times w_{k,k}w_{m,m}\rho_{l,n}+\frac{\delta_{i,j}}{d}
\end{align}
where $U\sim\mu_{\rm H}$ indicates that $U$ is distributed according to the Haar measure. The occurring expectation values can be obtained from Ref.~\cite{unexp}. We first consider the off-diagonal elements, i.e., the ones with $i\neq j$. Here, the expectation value vanishes unless $l=i$ and $n=j$, in which case we get
\begin{multline}
	\label{eq:obstemp1}
	\E_{U\sim\mu_{\rm H}}\left[U_{k,i}U_{m,j}\bar{U}_{k,j}\bar{U}_{m,i}\right]=\\
	\frac{1}{d(d^2-1)}\left(\delta_{k,m}(n-1)+(1-\delta_{k,m})(-1)\right).
\end{multline}
We now turn to the diagonal elements and note that if $i=j$, one needs $l=n$ to get a non-vanishing expectation value. We consider two cases separately: If $m=k$ we get
\begin{multline}
	\label{eq:obstemp2}
	\E_{U\sim\mu_{\rm H}}\left[U_{k,l}U_{k,i}\bar{U}_{k,l}\bar{U}_{k,i}\right]=\\
	\frac{1}{d(d^2-1)}\left(\delta_{i,l}2(n-1)+(1-\delta_{i,l})(n-1)\right),
\end{multline}
while for $m\neq k$ we obtain
\begin{multline}
	\label{eq:obstemp3}
	\E_{U\sim_\mu{\rm H}}\left[U_{k,l}U_{m,i}\bar{U}_{k,l}\bar{U}_{m,i}\right]=\\
	\frac{1}{d(d^2-1)}\left(\delta_{i,l}(n-1)+(1-\delta_{i,l})n\right).
\end{multline}
Inserting now (\ref{eq:obstemp1}), (\ref{eq:obstemp2}), and (\ref{eq:obstemp3}) into (\ref{eq:obstemp0}) and using $\sum_i\rho_{i,i}=1$, $\sum_iw_{i,i}=0$, and $\sum_i|w_{i,i}|^2=1$, we get $\mathcal{W}_{d,\rm H}(\rho)=\rho$ which concludes the proof.
\proofend

\subsection{Tight frames under physical restrictions}
In many situations of interest, the quantum state is not completely arbitrary but satisfies some additional conditions. In experiments with ultra-cold atoms, for example, the total particle number is conserved and super-selection rules forbid superpositions of states belonging to different eigenvalues of the corresponding operator.
Assume that the quantum state $\rho$ acts only in some subspace and denote the corresponding projection super-operator as $\mathcal{P}_N$. An important example is given by eigenspaces of an operator $\hat{N}$ describing a conserved quantity, i.e.\ one with commuting with $\rho$. In this case of a quantum state confined to a subspace, Definition \ref{def:tightframe} can be relaxed to
\begin{equation}
\label{eq:restrictedtf}
\mathcal{W}_{d_N}\circ\mathcal{P}_N=\mathcal{P}_N
\end{equation}
where $d_N$ denotes the dimension of the subspace corresponding to the eigenvalue $N$ of $\hat{N}$. If one aims at realizing such a restricted tight frame by the means of Theorem \ref{obs:inducedtf}, one can replace the group $\SU(d)$ by
\begin{equation}
\label{eq:surest}
\SU_{\hat{N}}(d)=\left\{U\in\SU(d):[U,\hat{N}]=0\right\}.
\end{equation}
As this group is compact, there exists a unique Haar probability measure on it. We adapt Definition \ref{def:inducedobs} to this situation.
\begin{definition}[Induced observable on subspace]
\label{def:inducedobssub}
Let $\hat{N}\in\mathcal{B}(\mathcal{H})$ and $S_N$ its $d_N$ dimensional eigenspace corresponding to the eigenvalue $N$. To a measure $\tilde{\mu}$ on the group $\SU_{\hat{N}}(d)$ and an observable $w_0\in S_N$ with $\|w_0\|_2=1$, we associate the following measure on $S_N$:
\begin{equation}
\mu=\biggl(1-\frac{1}{d_N^2}\biggr)\tilde{\mu}\circ f^{-1}+\frac{1}{d_N^2}\delta_{\id/\sqrt{d_N}}
\end{equation} 
where $f(U)=U^\dag w_0 U$.
\end{definition}
The matrices $\rho$, $w_0$, and all $V\in\SU_{\hat{N}}(d)$ are block diagonal with respect to the eigenbasis of $\hat{N}$. We consider the block corresponding to the eigenvalue $N$. Invariance of the Haar measure on $\SU_{\hat{N}}$ implies that if $V\sim\mu_{{\rm H},\SU_{\hat{N}(d)}}$, then $\mathcal{P}_N(V)\sim\mu_{{\rm H},\SU(d_N)}$. Thus, one can apply the proof of Theorem \ref{obs:inducedtf} to this block and obtain Eq.\ (\ref{eq:restrictedtf}).

\section{Unitary $t$-designs}
\label{sec:undesign}
The effort to implement random unitaries drawn from the Haar measure scales exponentially in the number of lattice sites $k$, making an implementation both theoretically inefficient and pratically unfeasible. However, this problem can be circumvented by replacing the Haar measure by a unitary $2$-design which is much easier to sample from as we will see later on. Unitary $t$-designs behave like the Haar measure in specific situations \cite{unidesign1,unidesign2,lowthesis}. The definition most suited to our problem is the following:
Let $\nu$ be a probability measure on $\SU(d)$. We define two channels on $\mathcal{B}(\mathcal{H}^{\otimes t})$:
\begin{equation}
\label{eq:Gnu}
\mathcal{G}_{t,\nu}(\rho)=\E_{U\sim\nu}\left[U^{\otimes t}\rho(U^\dag)^{\otimes t}\right]
\end{equation}
and $\mathcal{G}_{t,\mu_{\rm H}}$, where $\mu_{\rm H}$ is the Haar measure. We say that $\nu$ is a unitary $t$-design if $\mathcal{G}_\nu=\mathcal{G}_{\rm H}$. We say that $\nu$ is an $\eps$-approximate $t$-design if
\begin{equation}
\label{eq:approxdef}
\|\mathcal{G}_{t,\nu}-\mathcal{G}_{t,\rm H}\|\le\eps
\end{equation}
where $\|\cdot\|$ denotes the superoperator $2\rightarrow 2$-norm which is defined as 
\begin{equation}
\label{eq:22normdef}
\|\mathcal{O}\|=\sup_{X,\|X\|_2=1}\|\mathcal{O}(X)\|_2.
\end{equation}
This superoperator norm is equal to the Schatten $\infty$-norm when the channel is seen as a mere linear operator acting on the real vector space of Hermitian matrices. In the remainder we only consider the case $t=2$ and drop the index $t$ for simplicity.

\subsection{Tight frames from unitary 2-designs}

For our purpose, i.e., replacing the Haar measure random unitary in Theorem \ref{obs:inducedtf}, we need an approximate unitary $2$-design. This is the case because when Haar measure induces a tight frame, c.f.\ Eq.\ (\ref{eq:obstemp0}), both $U$ and $U^\dag$ appear twice. 
\begin{theorem}[Tight frames induced by unitary 2-designs]
\label{obs:inducedtfdesign}
Let $w_0$ be a traceless, 2-norm normalized observable and  $\nu$ be an $\eps$-approximate unitary $2$-design. Then
the sampling operator $\mathcal{W}_{d,\nu}$ corresponding to the induced measure fulfills
\begin{equation}
\label{eq:Wdiff}
\|\mathcal{W}_{d,\nu}-\id\|\le\sqrt{d}(d^2-1)\eps.
\end{equation}
\end{theorem}
\newproof   
We have 
\begin{align}
\|\mathcal{W}_{d,\nu}-\id\|=&\|\mathcal{W}_{d,\nu}-\mathcal{W}_{d,\rm H}\|\nonumber\\
=&\sup_{X,\|X\|_2=1}\|\mathcal{W}_{d,\nu}(X)-\mathcal{W}_{d,\rm H} (X)\|_2.\label{eq:temp1}
\end{align}
We note that
	\begin{align}
	\mathcal{W}_{d,\nu}(X)=&(d^2-1)\E_{U\sim\nu}\left[(U^\dag w_0U,X)U^\dag w_0U\right]+\frac{\id}{d}\Tr X\nonumber\\
	=&(d^2-1)\sqrt{d}\,\Tr_1\Biggl(\E_{U\sim\nu}\Bigl[(U^\dag\otimes U^\dag)\nonumber\\
	&\times(w_0\otimes w_0)(U\otimes U)\Bigr]\biggl(X\otimes\frac{\id}{\sqrt{d}}\biggr)\Biggr)
	+\frac{\id}{d}\Tr X,\label{eq:Wdesign}
\end{align}
where $\Tr_1$ denotes the partial trace with respect to the first of the two subsystems of equal dimension. Note that we have extended the definition of $\mathcal{W}_{d,\nu}$ to operators with non-unit trace. This relation yields (\ref{eq:Wdiff}) after inserting it into Eq.\ (\ref{eq:temp1}) and applying (\ref{eq:approxdef}).\proofend

The same argument holds also for restricted tight frames as defined in (\ref{eq:restrictedtf}): Let $\nu$ be a distribution on $\SU_{\hat{N}}(d)$ such that if $U\sim\nu$, then $\mathcal{P}_N(U)$ is drawn from an $\eps$-approximate 2-design on $\SU(d_N)$. This implies
\begin{equation}
\label{eq:apprestrictedtf}
\|\mathcal{W}_{d_N,\nu}\mathcal{P}_N-\mathcal{P}_N\|\le\sqrt{d_N}(d_N^2-1)\eps,
\end{equation}
as follows from the application of the above proof to the block in $U$ corresponding to the eigenvalue $N$ of $\hat{N}$. 

\subsection{Compressed sensing}

The technique of compressed sensing allows to reduce the number of measurements which are needed to reconstruct a quantum state from $\Theta(d^2)$ to $\Theta(d\,{\rm polylog}(d))$ if the rank of the state does not increase with $d$. To perform this method, one has to choose $m=\Theta(d\,{\rm polylog}(d))$ observables $w_1,\ldots,w_m$ randomly from the tight frame according to the corresponding probability measure and determine their expectation value $(w_i,\rho)$ by measurement. Then, one can efficiently solve the optimization problem
\begin{equation}
\label{eq:cs}
\min_\sigma\|\sigma\|_1\,{\rm s.t.}\,\forall i=1,\ldots,m:\,\,(w_i,\sigma)=(w_i,\rho).
\end{equation}
The theory has been developed for observables forming operator bases in Refs.\ 
\cite{davidletter,david} and extended to tight frames in Ref.\ \cite{cvcs}. There, it was also shown that the tight frame condition may be violated and compressed sensing is still possible if
\begin{equation}
	\label{eq:defalmosttf}
	\|\mathcal{W}_d-\id\|\le\frac{1}{8\sqrt{r}},
\end{equation}
where $r$ is the rank of the state we want to reconstruct. Not all tight frames are equally suited for compressed sensing as can be seen with a simple example: Let $w_1,\ldots,w_{d^2}$ be the elements of an orthonomal operator bases of $\mathcal{B}(\mathbb{C}^d)$ where $w_1$ is a rank-one projector and $\rho=w_1$. In this case, one has to measure of the order of $d^2$ observables before one ``hits'' $w_1$ and gets any information on the system. In Refs.\ \cite{david,davidletter,cvcs}, it has been shown that this problem cannot occur if all observables fulfill the so-called ``Fourier type incoherence condition'' which reads
\begin{equation}
\label{eq:fourier}
\PP_{w\sim\mu}\left(\|w\|_\infty^2 > \frac{\lambda}{d}\right)=0,
\end{equation}
where $\lambda$ must fulfill $\lambda=O({\rm polylog}(d))$. Note that statements like (\ref{eq:fourier}) make only sense when considering families of tight frames with growing dimension $d$. As we are mainly interested in the asymptotic efficiency of our scheme, we restrict ourselves to the scaling behavior and omit explicit prefactors. We now give a condition under which Eq.\ (\ref{eq:fourier}) is fulfilled in the situation of interest.

\begin{theorem}[Compressed sensing with induced observables]
\label{obs:inducedcs}
Let $w_0$ be a traceless, normalized observable fulfilling $\|w_0\|_\infty^2\le\lambda/d$ with $\lambda=O({\rm polylog}(d))$, and let $\nu$ be a $1/(8\sqrt{rd}(d^2-1))$-approximate $2$-design. The induced tight frame fulfills (\ref{eq:fourier}), which implies that it allows for compressed sensing.
\end{theorem}

\newproof Since 
\begin{equation}
	\|U^\dag w_0U\|_\infty^2=\|w_0\|_\infty^2 
\end{equation}
and $\|\id/\sqrt{d}\|_\infty^2=1/d$, condition (\ref{eq:fourier}) is fulfilled and from (\ref{eq:Wdiff}) it follows that (\ref{eq:defalmosttf}) is satisfied which proves that compressed sensing is possible.\proofend 

Theorem \ref{obs:inducedcs} holds especially in the important case of a observable which acts non-trivially only on a few number of lattice sites because here it is of the form
\begin{equation}
\label{eq:wtilde}
w_0=v\otimes\frac{\id_{d_l^{k-m}}}{({d_l^{k-m}})^{1/2}}
\end{equation}
for some normalized, traceless observable $v$ and some small constant $m$. 

\section{Approximation of unitary $2$-designs by random quantum circuits}
\label{sec:rand}
In the previous section, we have shown that unitary $2$-designs can be used to realize tight frames. We now show how they can be approximated by parallel random circuits and generalize the results of Refs.~\cite{randomcircuits,expspeedups,randomcircuits2} to show the following:
\begin{theorem}[Random circuits]
\label{obs:randomcircuits}
Assume $k$ to be even. Consider a parallel random circuit where in each step either $U_{1,2}\otimes U_{3,4}\otimes\ldots\otimes U_{k-1,k}$ or $U_{2,3}\otimes U_{4,5}\otimes\ldots\otimes U_{k-2,k-1}$ is performed with probability $1/2$ where $U_{i,i+1}$ acts on the neighboring sites $i$ and $i+1$. If the nearest-neighbor unitaries are drawn, in each step independently, from a probability measure $\nu_2$ which is universal, as defined below, there exists a constant $C$ (depending on the local dimension) such that the random circuit forms an $\eps$-approximate unitary $2$-design after $n=C\log(1/\eps)k\log k$ steps.
\end{theorem}
A finite set of nearest-neighbor unitary quantum gates is called universal if they generate a dense subgroup of $\SU(d_l^2)$. For an  arbitrary probability measure on $\SU(d_l^2)$, the notion of universality can be generalized, according to Ref.~\cite{randomcircuits}:
\begin{definition}[Universality]
\label{def:universality}
We say that $\mu$ is universal if for any open ball $S$ there exists $l>0$ such that $S$ has a nonzero weight for the $l$-fold convolution product of $\mu$.
\end{definition}

\newproof The proof of Theorem \ref{obs:randomcircuits} is an extension of that of similar results in Refs.\ \cite{expspeedups,randomcircuits2}. Readers mostly interested in the application to optical lattice systems can safely skip to the next section.

In Ref.\ \cite{randomcircuits2}, it is shown that parallel random circuits with periodic boundary conditions form $\eps$-approximate $2$-designs after $n=C\log(1/\eps)k$ steps if the unitaries are drawn from the Haar measure on $\SU(d_l^2)$. 

We now proceed in three steps: First, we show that the nearest-neighbor unitaries can be drawn from an approximate $2$-design on $\SU(d_l^2)$ instead from the Haar measure. We denote the measure corresponding to a single step of the random circuit by $\nu_k$ and define the linear operator $G_{\nu_k}$ by
\begin{equation}
\label{eq:Gnuk}
G_{\nu_k}=\int{\rm d}\nu_k(U)\,U^{\otimes 2}\otimes\bar{U}^{\otimes 2}.
\end{equation}
This operator can be decomposed as $G_{\nu_k}=(M_e+M_o)/2$ with 
\begin{align}
M_e&=P_{1,2}\otimes P_{3,4}\otimes\ldots\otimes P_{k-1,k},\\
M_o&=P_{k,1}\otimes P_{2,3}\otimes\ldots\otimes P_{k-2,k-1}\label{eq:Mo}
\end{align}
where
\begin{equation}
P_{i,j}=\int{\rm d}\nu_2(U_{i,j})\,U_{i,j}^{\otimes 2}\otimes\bar{U}_{i,j}^{\otimes 2}.
\end{equation}
We have to bound 
\begin{equation}
	\|G_{\nu_k}^n-G_{{\rm H}}\|_\infty=\lambda_2(G_{\nu_k})^n
\end{equation}	
where $n$ is the depth of the circuit and $\lambda_2$ denotes the second largest eigenvalue. In Ref.~\cite{randomcircuits2}, it is shown that if the nearest-neighbor unitaries are drawn from the Haar measure, there exists a constant $\Lambda>0$ such that the corresponding operator $\tilde{G}_{\nu_k}$ fulfills $\lambda_2(\tilde{G}_{\nu_k})\le 1-\Lambda$. Using now the fact, that $\nu_2$ is a $\delta$-approximate $2$-design, we get
\begin{equation}
\lambda_2(G_{\nu_k})\le\lambda_2(\tilde{G}_{\nu_k})+k\|G_{\nu_2}-G_{{\rm H}_2}\|_\infty\le 1+k\delta-\Lambda.
\end{equation}

For the right-hand side to be smaller than one, which is necessary and sufficient for an exponentially fast convergence, one has to choose  $\delta=O(1/k)$. To realize this $\textit{local}$ approximate $2$-design, we use a result from 
Ref.\ \cite{randomcircuits} which states that one needs to draw only $s=O(\log(1/\delta))$ gates from an arbitrary universal gate set 
to achieve this. We pick $s$ to be a power of $2$, which is surely always possible.
Thus, we have a circuit with depth 
\begin{equation}
\label{eq:ndepths}
n_\eps=O(ns)=O(\log(1/\eps)k\log k)
\end{equation}
where the random choice between $M_e$ and $M_o$ is not made in every step but in blocks of $s$ steps which corresponds to the operator $(M_e^s+M_o^s)/2$ while $s$ steps of the actual quantum circuit performed are described by $((M_e+M_o)/2)^s$. They both 
have the same fixed point. What is more, Theorem \ref{Mixing} holds true, see below.
This means, in particular, that 
the convergence of the actual circuit cannot be slower and (\ref{eq:ndepths}) holds.  

The last thing needed to obtain Theorem \ref{obs:randomcircuits} is to switch to open boundary conditions, i.e., remove the first tensor factor in Eq.\ (\ref{eq:Mo}). As this does not affect the fixed point and the operator norm difference is on the order of $\delta$, only the prefactor is changed slightly.  We note that the prefactor depends on the actual choice of $\nu_2$. 

The conditions for Theorem \ref{def:inducedobs} to apply are fulfilled if $1/\eps=O(d^{5/2})$.  Using this in 
Eq.\ (\ref{eq:ndepths}), we get
\begin{equation}
	\label{eq:totaldepths}
	n_{\rm T}=O(k^2\log k).
\end{equation}
This means compressed sensing is possible with a single traceless observable and a parallel random quantum circuit with a depth given by Eq.\ (\ref{eq:totaldepths}). 

We did not explicitly discuss the case of restricted tight frames because it can be, again treated by block-decomposing all matrices according to the spectral decomposition of the operator $\hat{N}$ describing the symmetry, c.f. Eq.~(\ref{eq:surest}). 
Thus, with a parallel random quantum circuit as in Theorem \ref{obs:randomcircuits} with unitaries which are universal for $\SU_{\hat{N}}(d_l^2)$ with a depth as in Eq.\ (\ref{eq:totaldepths}), one can perform compressed sensing for states within some eigenspace of $\hat{N}$.\proofend

\begin{theorem}[Mixing properties of circuits]\label{Mixing}  Let $s$ be a power of $2$, then
\begin{equation}
\label{eq:fastermixing}
\lambda_2\left(\left(\frac{M_e^2+M^2_o}{2}\right)^s\right)\le\lambda_2\left(\frac{M_e^{2s}+M_o^{2s}}{2}\right),
\end{equation}
\end{theorem}

\newproof We now prove the validity of Eq.\ (\ref{eq:fastermixing}) for $s$ being a power of $2$, i.e.\ $s=2^j$. Let $A$ and $B$ be two Hermitian matrices. From
$0\le(A-B)^2$ it follows that
\begin{equation}
	\left(\frac{A+B}{2}\right)^2\le\frac{A^2+B^2}{2}
\end{equation}
and hence also, from the monotonicity of eigenvalues \cite{Bhatia},
\begin{equation}	
\lambda_2\left(\left(\frac{A+B}{2}\right)^2\right)\le\lambda_2\left(\frac{A^2+B^2}{2}\right).\label{eq:genAB}
\end{equation}
We also need the fact that, for all pairs of positive Hermitian matrices $A$ and $B$,
\begin{equation}
\label{eq:fastermixing1}
\lambda_2(B)\ge\lambda_2(A)\Rightarrow\lambda_2(B^i)\ge\lambda_2(A^i),
\end{equation}
which can be easily seen by diagonalizing $A$ and $B$. 
Note that $M_e^2$ and $M_o^2$ are  positive operators. 
For any natural $i_1$, $i_2$ one finds that 
\begin{align}
\lambda_2\left[\left(\frac{M_e^{2^{i_1}}+M_o^{2^{i_1}}}{2}\right)^{2^{i_2}}\right]=&\lambda_2\left[\left(\left(\frac{M_e^{2^{i_1}}+M_o^{2^{i_1}}}{2}\right)^2\right)^{2^{i_2-1}}\right]\nonumber\\
\le&\lambda_2\left[\left(\frac{M_e^{2^{i_1+1}}+M_o^{2^{i_1+1}} }{2}\right)^{2^{i_2-1}}\right],
\label{eq:fastermixing2}
\end{align}
where we have employed (\ref{eq:fastermixing1}) and (\ref{eq:genAB}) for $A=M_e^{2^{i_1}}$ and $B=M_o^{2^{i_1}}$.
By a repeated application of (\ref{eq:fastermixing2}), starting from $i_1=1$ and $i_2=j$, until $i_1=j$ and $i_2=j$,
we finally get 
\begin{equation}
\lambda_2\left(\left(\frac{M_e^2+M_o^2}{2}\right)^{2^j}\right)\le\lambda_2\left(\frac{M_o^{2^{j+1}}+M_e^{2^{j+1}}}{2}\right),
\end{equation}
which is the statement to be shown.
\proofend

\section{Optical lattice systems}
\label{sec:optlat}

We now introduce a novel method for quantum state tomography for systems of ultra-cold atoms in optical lattices that does
{\it not} require any local addressing of single sites. In fact, each of the steps necessary has already been demonstrated experimentally.
The idea is to make use of appropriate randomness and natural time evolution, suitable exploiting 
optical superlattices, such that 
time-of-flight images give rise to complete tomographic information -- quite an interesting and promising
perspective. Using these experimental
tools, the above mathematical methods become applicable. It should be clear, as pointed
out before, that such a prescription also allows for detecting entanglement entropies \cite{Demler,Area} in systems of cold atoms.

Ultra-cold atoms in optical lattices form some of the cleanest quantum many-particle systems available for experiments and allow for the realization of well-known effects from condensed matter. For example, both the bosonic super-fluid Mott-insulator transition and the Mott state of fermions where observed by changing the intensity of the laser forming the lattice \cite{superfluidmott,fermimott}. Such systems also have the potential of functioning as quantum simulators which means, they allow to simulate systems from other branches of physics like the notoriously difficult quantum chromodynamics (QCD) \cite{quantumsimulators}.

\subsection{Time-of-flight measurements}

Even though measurements with spatial resolution have been demonstrated in recent experiments \cite{superlatticesrelax}, the standard technique is still provided by time-of-flight absorption imaging. Here, the lattice and the confining trap are instantaneously switched off and, after some time during which the atoms expand approximately without interaction, an absorption image is taken \cite{superfluidmott,tof}. As the distance that the atoms fly during the expansion is proportional to the initial momentum, this procedure amounts to a measurement of the density in momentum space. We restrict ourselves to the bosonic case, while noting that fermions can be treated in a completely analogous way, and expand the field operators of the one-dimensional bosonic field as
\begin{equation}
\label{eq:Wanndecomp}
\hat{\Psi}(x)=\sum_{j=1}^\infty\sum_{s=1}^kW^{(j)}(x-x_s)\hat{b}_s^{(j)}
\end{equation}
where $W^{(j)}$ is the Wannier function of the $j$-th band, $x_s$ is the position, and $\hat{b}_s^{(j)}$ the corresponding annihilation operator at site $s$. If the lattice is sufficiently deep, all bands but the lowest one can be neglected, and we drop the upper index in (\ref{eq:Wanndecomp}). The momentum-space distribution, which is measured in the time-of-flight experiment, is given by $n(p)=|\tilde{W}(p)|^2S(p)$ where $\tilde{W}$ is the Fourier transform of the Wannier function and the quasi-momentum distribution is given by 
\begin{equation}
\label{eq:quasimoment}
	S(p)=\sum_{s,l=1}^ke^{ip(x_s-x_l)}\langle\hat{b}_s^\dag\hat{b}_l\rangle.
\end{equation}
As we do not assume the state to be translationally invariant, Eq.~\ (\ref{eq:quasimoment}) cannot be inverted to get the two-point correlation functions in real space but we can get for integer $l$
\begin{equation}
	\label{eq:correlator}
	\sum_{s=1}^k\langle\hat{b}_s^\dag\hat{b}_{s+l}\rangle=\frac{1}{2\pi}\int_{-\pi}^\pi{\rm d}p\,e^{ipl}S(p),
\end{equation}
where we have set the lattice spacing to one. If the atoms are bosons, the local Hilbert space is infinite-dimensional. However, as the interaction between the atoms must always be repulsive to ensure stability of the quantum gas, one can neglect state where more than a given cut-off number of atoms are present on a single lattice site. This allows us to work with a finite-dimensional Hilbert space. We note that this does not even needs to be an approximation as one can set the maximal number of bosons per site $N_S$ to their total number $N$. However, in any practical setting, one would use $N_S\ll N$ and still get a very good approximation. 
For an arbitrary $i$, we set 
\begin{equation}
	\label{eq:w0b}
	w^{(i)}_0\propto\sum_{j=1}^k(\hat{b}_i^\dag\hat{b}_{i+j}+\hat{b}_{i+j}^\dag\hat{b}_j)
\end{equation}
which is traceless. Due to the sum, which stems from the absence of translational invariance, Eq.\ (\ref{eq:w0b}) is not exactly of the form given by (\ref{eq:wtilde}) but a sum of few, i.e.\ logarithmically many in the Hilbert space dimension $d$, terms of this form. This implies
\begin{equation}
\|w_0^{(i)}\|_\infty\le C\sum_{j=1}^k\|\hat{b}^\dag_j\hat{b}_{i+j}+\hat{b}_{i+j}^\dag\hat{b}_j\|_\infty\le\tilde{C}k
\end{equation}  
where $C,\tilde{C}>0$ are constants. Because $k=\Theta(\log d)$, we can employ Theorem \ref{obs:inducedtf} to show that a measurement of the momentum space distribution, together with an approximate $2$-design, allows for efficient compressed sensing. 

Although already a single choice of $i$ yields an approximate tight frame, we can use the data corresponding to all $i=1,\ldots,k$, as they are measured anyway, to reduce the necessary number of experiments.

\subsection{Realization of the random circuit}
We now discuss how a probability measure on the nearest-neighbor unitaries that is universal can be obtained. To be as specific and simple as possible, we use the single-band Bose Hubbard model with Hamiltonian \cite{bhm}
\begin{equation}
\label{eq:H}
\hat{H}=- \sum_{i=1}^k J_i(\hat{b}_i^\dag\hat{b}_{i+1}+\hat{b}_{i+1}^\dag\hat{b}_i)+\frac{U_i}{2}\hat{n}_i(\hat{n}_i-1)+\Delta_i\hat{n}_i,
\end{equation}
where $\hat{n}_i=\hat{b}_i^\dag\hat{b}_i$ and $J_1\dots, J_k;\Delta_1,\dots,\Delta_k;U_1,\dots, U_k\in \rr$.
To realize the parallel random quantum circuit, we make use of the techniques of super-lattices \cite{superlattices,superlatticesrelax,superlatticetheory} 
and speckle patterns \cite{speckle}. As the total number of atoms is conserved and super-selection rules forbid the superposition of states with different particle numbers, we restrict ourselves to particle-number conserving operations. 

By using an additional lattice for which the lattice constant is twice as large, one can change the height of the wells between alternating pairs of site. This mainly affects the hopping constants $J_i$ and to much less extent the interaction parameter $U_i$, an effect which we neglect. By choosing the depth of the super-lattice large enough, we get $J_i=0$ for the non-coupled pairs and $J_i=J$ for the coupled ones. Such a double-well structure has been used in Ref.\ \cite{superlatticesrelax} 
to probe correlation functions. A speckle pattern is created by illuminating an uneven surface with a laser and can be modelled by a spatially fluctuating $\Delta_i$. The situation is sketched in Figure \ref{fig:optlat}. For reasons of simplicity, we only consider the regime for which the strength of the speckle potential is not correlated over the lattice sites, i.e., all $\Delta_i$ are independently distributed. Thus, we have a random circuit as in Theorem \ref{obs:randomcircuits} and we only need to show that the corresponding gates generate a dense set in $\SU_N(d_l^2)$. The local gates are
\begin{align}
	\label{eq:U}
	U_{i,j}(\Delta_1,\Delta_2,t)=&\exp\bigl(-it(-J(\hat{b}_1^\dag\hat{b}_2+\hat{b}_2^\dag\hat{b}_1)\nonumber\\
	&+\frac{U}{2}(\hat{n}_1(\hat{n}_1-1)+\hat{n}_2(\hat{n}_2-1))\nonumber\\
	&+\Delta_1\hat{n}_1+\Delta_2\hat{n}_2)\bigr),
\end{align}
where $t\geq 0$ is the time after which the super-lattice is switched and a new realization of the laser speckle is created. We assume $\Delta_{1,2}$ to be Gaussian distributed, and we have neglected global phases.  Now one can adopt an argument used in Ref.~\cite{cvcomputing} for showing that the Gaussian operations together with a single non-Gaussian one allow for continuous-variable quantum computation. For universality to hold, one has to generate all operations where the corresponding Hamiltonian is a polynomial in the creation and annihilation operators where every monomial must contain an equal number of creation and annihilation operators, i.e., must be balanced, to ensure particle-number conservation. This is true as Eq.\ (\ref{eq:U}) contains all quadratic terms and a single 
quartic one. Since one can generate the entire 
algebra generated from the original set of Hamiltonian by commutation, one can approximate an 
arbitrary unitary \cite{cvcomputing}. 
Thus, by varying $\Delta_{1,2}$, we can approximate any gate to arbitrary accuracy which implies, by continuity of Eq.\ 
(\ref{eq:U}) universality as in Definition \ref{def:universality}. By appropriately choosing the distribution from which $t$ is chosen, the set can be made closed under Hermitian conjugation. Now, we can apply Theorem \ref{obs:randomcircuits} which shows together with Theorem \ref{obs:inducedtf} that one can perform efficient compressed sensing by using a optical super-lattices, laser speckles, and time-of-flight imaging.  
Again, no single site addressing is necessary, and still complete tomographic information is obtained.

\section{More efficient tomography scheme for matrix product states and operators}

\label{sec:mps}
Even though compressed sensing notably reduces the number of necessary measurements, it still scales exponentially with the number of lattice sites. Without any further assumption on the state, this cannot be overcome. However, when the state is described by a generic matrix-product state (MPS) with a fixed bond dimension, tomography is possible with the number of measurements growing, in general, almost linearly with the system size and being constant for translationally invariant MPS. Ground states of gapped Hamiltonians which are a sum of terms acting only on a constant number of lattice sites are, generically, of this type \cite{mpstheory}.

\subsection{Reconstructing reduced density matrices}

The exponential reduction of the necessary number of measurements if the state is a MPS is due to the fact that such states are completely determined by their reduced density matrices on all blocks of $l$ consecutive sites where $l$ only depends on the bond dimension \cite{mpstheory,mps}. In Ref.~\cite{mps}, an efficient algorithm is given for finding the MPS matrices from these reduced density matrices. Note that this procedure only works if an upper bound to the bond-dimension, or, equivalently, to the locality size of the Hamiltonian is known. This is obvious as there could always be long range correlations which do not affect the $l$ site reduced density matrices.

Let $d_B=d_l^l$ be the dimension of the subsystem under consideration and define the operator $\mathcal{T}_{R_q}:\mathcal{B}(\mathbbm{C}^d)\mapsto\mathcal{B}(\mathbbm{C}^{d_B})$ be the operator acting as $\mathcal{T}_{R_q}(\rho)=\Tr_{R_q}\rho$ and $R_q$ denote all lattice sites but $q,\ldots,q+l-1$. As we only want to perform tomography on the lattice sites $q,\ldots,q+l-1$, we can trace out the remainder of the system. In this case, the tight-frame condition of Definition \ref{def:tightframe} becomes
\begin{equation}
\label{eq:reducedtf}
\mathcal{T}_{R_q}\circ\mathcal{W}_{d_B}=\mathcal{T}_{R_q}. 
\end{equation}
If additional constraints apply, the condition reads
\begin{equation}
\label{eq:reducedrestrictedtf}
\mathcal{T}_{R_q}\circ\mathcal{W}_{d_{BN}}\circ\mathcal{P}_N=\mathcal{T}_{R_q}\circ\mathcal{P}_N
\end{equation}
where $d_{BN}$ is the dimension of the matrix block corresponding to the eigenvalue $N$ of $\hat{N}$, restricted to the subsystem of $l$ lattice sites. 
We concentrate on the former case as the second follows in an analogous way and show that a random circuit with a depth which does not depend on the system size can realize such a reduced tight frame: Assume $w_0$ to be a sum of traceless observables, each acting non-trivially only on some block of $l$ lattice sites. For compressed sensing to be possible, it is sufficient for $\nu$ to induce, for every $q$, an approximate reduced tight frame with 
\begin{equation}
\label{eq:appredtf}
	\|\mathcal{T}_{R_q}(\mathcal{W}_{d_B,\nu}-\id)\|\le\sqrt{d_B}(d_B^2-1)\eps
\end{equation}
where $\eps$ must be chosen such that $\sqrt{d_B}(d_B^2-1)\eps<1/(8\sqrt{r}$, c.f.\ Eq.\ (\ref{eq:defalmosttf}). This is the case if it is an approximate reduced unitary $2$-design with
\begin{equation}
\label{eq:appredtfrestricted}
\forall q:\sup_{X,\|X\|_2=1}\biggl\|\Tr_{R_q}(\mathcal{G}_\nu-\mathcal{G}^{(q)}_{{\rm H}_l})(X)\biggr\|_2\le\eps.
\end{equation}
where $\mathcal{G}^{(q)}_{{\rm H}_l}$ denotes the channel which acts as $\mathcal{G}_{\rm H}$ on a block of $l$ sites starting at $q$ and as the identity on the rest of the system. To see that this is true, we calculate
\begin{multline}
\|\mathcal{T}_{R_q}(\mathcal{W}_{d_B,\nu}-\id)\|=\\
\sup_{X,\|X\|_2=1}\|\Tr_{R_q}(\mathcal{W}_{d_B,\nu}(X)-\mathcal{W}_{d_B,\rm H}(X))\|_2,
\end{multline}
which yields the desired result after inserting (\ref{eq:Wdesign}) and applying (\ref{eq:appredtfrestricted}).

To show that obtaining such a tight frame is efficiently possible, we adapt Theorem \ref{obs:randomcircuits}: 
\begin{theorem}[Reduced tight frames by reduced unitary $2$-designs]
\label{obs:reduceddesign}
Let the parallel random circuit be as in Theorem \ref{obs:randomcircuits}. There exists some constant $C$ such that it forms an 
$\eps$-approximate reduced $l$-site $2$-design as defined in (\ref{eq:appredtf}) after $n=C\log(1/\eps)l\log l$ steps. 
\end{theorem}

\begin{figure}
\begin{center}
 \includegraphics[width=0.7\linewidth]{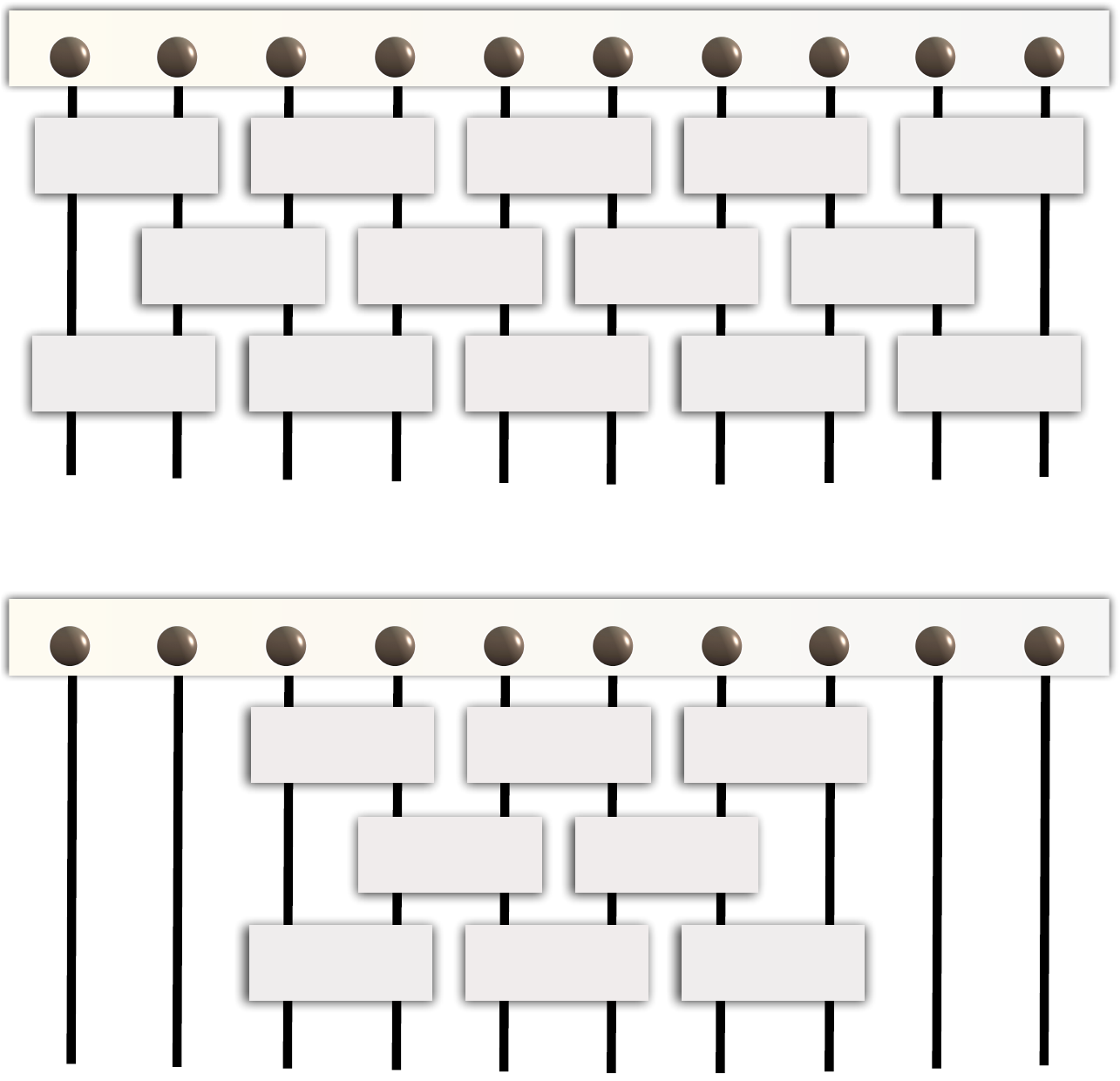}
\caption{\label{fig:comparecircuits}Top: Parallel random circuit acting on the entire system. Bottom: Random circuit used in the proof of Theorem \ref{obs:reduceddesign} which acts only on the inner sites. The above circuit randomizes the states of the inner sites not less than the above one.}
\end{center}
\end{figure}

\newproof We show the fast convergence of our random circuit to a reduced unitary $2$-design by comparing it with another random circuit which is easier to deal with, see Fig.~\ref{fig:comparecircuits}. In analogy to Eq.\ (\ref{eq:Gnuk}) we denote by $\mathcal{G}^{(q)}_{\nu_l}$ the channel corresponding to an application of the  parallel quantum circuit to a block consisting of $l$ lattice sites starting from $q$. Theorem \ref{obs:randomcircuits} implies 
\begin{equation}
\sup_{X,\|X\|_2=1}\|\Tr_{R_q}((\mathcal{G}^{(q)}_{\nu_l})^n-\mathcal{G}^{(q)}_{{\rm H}_l})(X)\|_2\le\eps
\end{equation}
for $n=C\log(1/\eps)l\log l$ where $C$ is a constant. Using that $\lambda_2(G_{\nu_k})\le \lambda_2(G_{\nu_l})$, one obtains (\ref{eq:appredtfrestricted}), concluding the proof. \proofend

\subsection{Complexity of classical post-processing}

\begin{figure}
\begin{center}
 \includegraphics[width=0.7\linewidth]{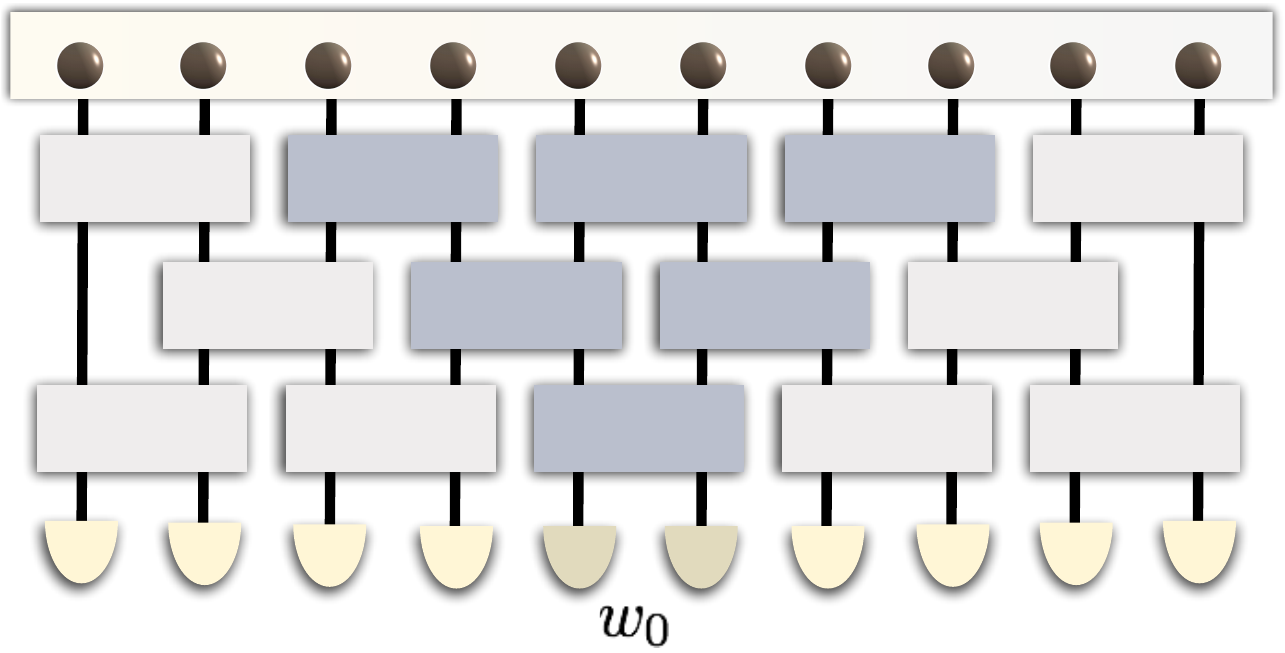}
\caption{\label{fig:causcone}Influence region for a local observables. Only the action of the darkly colored gates influences the measured observables.}
\end{center}
\end{figure}

Theorem \ref{obs:reduceddesign} implies that a random quantum circuit of constant depth is sufficient to perform tomography on a reduced density matrix of constant size. The number of operations, i.e.\ random unitary gates and measurements of expectation values of $w_0$, does only scale polynomially with the number of lattice sites. Therefore, we regard the quantum part of the protocol as efficient. However, this says nothing about the amount of post-processing needed because the reconstruction of the $l$-site reduced density matrices requires the knowledge of $\Tr_{R_q}w$ for all observables $w$ obtained by the realizations of the random quantum circuit. If one needed to keep track of the evolution of observables on the entire 
Hilbert space this would require an exponential amount of computational resources. 
To see that this is not a problem in the present situation, we use the fact that 
$w_0$ is a sum of terms which act non-trivially only on blocks of constant size. Thus, the observables induced by the constant-depth random circuit can be written as 
\begin{equation}
\label{eq:tttemp1}
	w(U)=\sum_i w_i(U)\otimes\id_{R_i}/({d_{R_i}})^{1/2}
\end{equation}
where $w_i(U)$ acts non-trivially only on a block of constant size $L$ starting with site $i$ and where $R_i$ denote all lattice sites but $i,\ldots,i+L-1$, see Figure \ref{fig:causcone}. From Eq.\ (\ref{eq:tttemp1}), we get
\begin{equation}
	\label{eq:wU}
	\Tr_{R_q} (w(U))=\sum_i\sqrt{d_{R_i}}\Tr_{R_q} (w_i(U)),
\end{equation}
which means that one only needs to deal with observables on $k$ Hilbert spaces which all have dimension $d_l^L$ not depending on $k$ making also the classical part of the protocol efficient. If the system is assume to be translationally invariant, all reduced density-matrices are equal, reducing the necessary number of measurements to a constant. Roughly speaking, the random circuit transforms a local observable to a reduced tight frame on $l$ lattice sites with some influence on $L>l$ sites and none on the rest of the system. 

\subsection{Mixed states}

Even though the method developed in this section is, in its 
present form, limited to pure states, it can be naturally extented to mixed states by using recent results on tomography for matrix-product \textit{operators} (MPO) which are a natural generalization of MPS \cite{mpotheory1, mpotheory2}. In Ref.\ \cite{mpo}, it is shown that large classes of MPO states can be 
efficiently reconstructed from the reduced density matrices on a constant number of sites not depending on the system size. As we have presented a method of recovering  these objects, one directly obtains a way of performing tomography on states in optical lattices which are described be MPO while requiring the same experimental techniques.

\section{Conclusion}

\label{sec:conc}
In this article, we have presented a new route towards efficient quantum state tomography for quantum many-body systems, specifically for bosons in optical lattices. By using random circuits, which can be implemented by means of super-lattices and laser speckles, 
one can avoid the use of tomographically complete local measurements on single sites (in fact any local addressing) -- still complete
tomographic knowledge can be achieved.
These are challenging, needless to say, but rely solely on time-of-flight imaging techniques, which are nowadays
routinely implemented. Without any further assumptions to the state, the number of necessary measurements is optimal up to constants and logarithmic factors in the systems dimension.
Restricting the set of possible states to matrix product states, both the number of measurements 
and the depth of the required random quantum circuit does not at all depend on the system size. This idea gives rise to the
exciting perspective of actually reiably measuring out the full quantum state of a quantum many-body system in the
laboratory.

There are a number of questions arising from this: For example, it would be very interesting to compare the performance of the present scheme, which is based on random circuits, with one where the quantum gates are chosen in an optimal way from some set of feasible operations. While there is not much room for improvements concerning the asymptotic behavior, the performance for small systems might differ notably. This will be studied extensively by numerical means in forthcoming work. We hope that this work stimulates further work -- both of theoretical and especially experimental kind -- in ``quantum system identification'', in order to innovate ways of 
rendering quantum state tomography feasible for large quantum systems.

\section{Acknowledgments.}

We thank M.\ Kliesch and C.\ Gogolin for valuable discussions, especially in the initial phase of this project. We would like to thank the EU (Qessence, Compas, Minos), the BMBF (QuOReP), and the EURYI for support.

\end{document}